\documentclass[superscriptaddress,secnumarabic,twocolumn,
 amssymb,amsmath,nobibnotes,aps,prd,showkeys,showpacs]{revtex4}
\usepackage{graphicx,epsfig,amssymb,amsmath,amstext}
\usepackage{bm}
\usepackage{booktabs}

\newcommand{\abs}[1]{\left| #1 \right|} % for absolute value
\renewcommand{\d}[2]{\frac{d #1}{d #2}} % for derivatives
\newcommand{\pd}[2]{\frac{\partial #1}{\partial #2}} 
\newcommand{\grad}[1]{\mathbf{\nabla} #1} % for gradient
\let\baraccent=\= % rename builtin command \= to \baraccent
\renewcommand{\=}[1]{\stackrel{#1}{=}} % for putting numbers above =

\begin{document}
\title{A Symmetric Integrator for non-integrable Hamiltonian Relativistic Systems}
\author{Jonathan Seyrich}
\affiliation{Mathematisches Institut, Universit\"{a}t T\"{u}bingen,
Auf der Morgenstelle, 72076 T\"{u}bingen, Germany}
\email{seyrich@na.uni-tuebingen.de}

\author{Georgios Lukes-Gerakopoulos}
\affiliation{Theoretical Physics Institute, University of Jena, 07743 Jena,
Germany}
\email{gglukes@gmail.com}
\begin{abstract}
 By combining a standard symmetric, symplectic integrator with a new step
 size controller, we provide an integration scheme that is symmetric, reversible
 and conserves the values of the constants of motion. This new scheme is
 appropriate for long term numerical integrations of geodesic orbits in spacetime
 backgrounds, whose corresponding Hamiltonian system is non-integrable, and,
 in general, for any non-integrable Hamiltonian system whose kinetic part depends
 on the position variables. We show by numerical examples that the new integrator
 is faster and more accurate i) than the standard symplectic integration schemes
 with or without standard adaptive step size controllers and ii) than an adaptive
 step Runge-Kutta scheme.   
\end{abstract}

\maketitle

\section{Introduction}\label{sec-I}

In a few years a new era of astronomy is expected to begin when the Advanced LIGO
and other gravitational wave detectors are anticipated to receive the first signals.
The main source of gravitational wave signals are binary systems consisting of
compact objects whose mass proportion ranges from equal masses to extreme mass
ratio. The gravitational wave signals from binary systems are closely related to
the motion of the inspiraling compact objects. Thus, orbital studies of binary
systems have attracted much attention (e.g., \cite{Levinetal,Hackmann}). In general,
orbital motion can be studied by using a Hamiltonian function which splits into an
orbital and a spin part.

One possible approximation is the post-Newtonian (e.g., \cite{PostNewton} and
references therein). In the post-Newtonian approximation the Hamiltonian's
orbital part is expanded in powers of $\frac 1 {c^2}$. This approach has also
been extended to binaries that are perturbed by a third, much lighter object
(e.g., \cite{3body}). These post-Newtonian approaches have in common that
the mass ratio of the binary should not be very  large.

When the binary system consists of compact objects of extreme mass ratio, 
the motion in such a system is called an Extreme Mass Ratio Inspiral (EMRI). We
expect to find such binaries at the center of galaxies, where a comparably light
compact object inspirals into a central Supermassive Black Hole (SMBH). The
lighter inspiraling compact object moves adiabatically from one geodesic orbit
to another. Thus, an EMRI can be approximated by the geodesic
motion of a test particle in the spacetime background of the central
supermassive object (e.g., \cite{Amaro12} and references therein).

One interesting aspect of EMRI is that the shape of the spacetime background is
encoded in the gravitational waves emitted by the inspiraling object. Ryan
showed that we can, in principle, extract this information from the
gravitational waves \cite{Ryan}. Therefore, when future low frequency
space-borne gravitational wave detectors are launched, we will be able to test
whether the spacetime backgrounds around SMBHs are indeed described by the Kerr
metric or by something else \cite{Amaro12}. In order to provide such a test,
Collins and Hughes \cite{Collins04} constructed a perturbed Schwarzschild
spacetime background and called it a \textit{bumpy black hole}. Since then,
several authors have suggested different tests \cite{Glampedakis06,Barausse07,
Gair08,Apostolatos09,Vigeland10a,Vigeland10b,Lukes10,JohanPsaltis,BambiBar,
Vigeland11,Gair11,Bambi,Johannsen12,Pappas12} using various forms of bumpy black
holes (e.g., \cite{ManNov92,MMS,MSM}), which are also known as non-Kerr metrics.

Usually, these non-Kerr metrics are stationary and axisymmetric, but they miss
the fourth integral which would make them integrable systems, as the Carter
constant \cite{Carter68} does in the case of the Kerr metric. Thus, in
general the non-Kerr metrics correspond to non-integrable systems, which in
practical calculations often are investigated for chaos (e.g., \cite{Lukes10,Wu,
Dubeibe07,Han08,Semerak,Lukes}). This, in turn, implies numerical simulations
over long time intervals. Therefore, high-efficient integrators with a good
long term behavior are required.

Over the last decades, part of the Numerical Analysis community has focused on
long-term integrations of differential equations. Geometric or structure
preserving algorithms, such as symplectic methods for Hamiltonian equations of
motions or symmetric integrators for reversible systems, have been developed
aplenty. Contrary to standard explicit integrators such as
Runge-Kutta schemes, these integrators nearly conserve first integrals (e.g.,
the energy) and overall integration errors increase only slowly with time.
Whereas for standard integration schemes the overall error is normally
proportional to the square of the length of the integration interval $t_i$, it
only increases linearly with $t_i$ for structure preserving integrators. A 
detailed discussion of such methods can be found in \cite{hairerlubichwanner}.
 
The efficiency of numerical integrators can be increased by the use of adaptive
step size controllers. But, by using standard step size controllers for
geometric algorithms, we lose the integrators' good long term behavior. However,
over the last 15 years, algorithms allowing structure preserving step size
control have been developed (\cite{stoffer,soederlind}). This development has
resulted in a lot of available tools to confront problems in classical astronomy.  

But, in general relativity, the structure of the equations is more complex.
The inclusion of spins to a binary system leads to a Poisson-system, which can
be transformed to symplectic form as stated by the \textit{Darboux-Lie-Theorem}.
Such an approach was proposed in \cite{wuxie}. After such a transformation,
symplectic  schemes can finally be applied, e.g. \cite{zhongwu}. But also for the
the post-Newtonian formulation as a Poisson-system, a non-canonically symplectic
integrator has been found, see \cite{lubich}. In the geodesic approach, the
Hamiltonian of the orbit has a position dependent mass-matrix. Thus, from this
point of view, the system of differential equations is non-separable. Furthermore,
the vector field of the resulting equations of motion is strongly varying in
non-integrable spacetime backgrounds. These two features pose a new set of problems
for the implementation of efficient and accurate algorithms. As an answer to these
problems we introduce a new integration scheme for such differential equations
based on Gauss-Runge-Kutta collocation methods. In particular, we use a new
structure preserving adaptive step size controller. As a result, we get a symmetric,
reversible integrator which efficiently conserves the constants of motion. We
tested this integrator's performance in the case of the Manko, Sanabria-G\'{o}mez,
Manko (MSM) metric \cite{MSM} and the results show that the new integration scheme
is much faster than an adaptive step fifth-order Cash-Karp Runge-Kutta scheme.

The paper is organized as follows: Section \ref{sec-II} summarizes some 
basic elements about the MSM metric, and the geodesic motion in the MSM spacetime
background. In the subsequent Section \ref{sec-III}, we consider the topic from
a Numerical Analysis point of view and discuss the characteristics of the
geodesic equations. Our step size controller, along with the integrator, is
presented in section \ref{sec-IV}. The performance of the integrator in
numerical tests is shown in section \ref{sec-V}. Section \ref{sec-VI} summarizes
the main results of the study.

\section{Manko, Sanabri-G\'{o}mez, Manko spacetime background} \label{sec-II}

 \subsection{The metric} \label{subsec-II-a}

 We tested the various integration schemes on the five-parameter vacuum solution
 introduced by Manko et al. \cite{MSM}. The MSM solution is asymptotically flat,
 axisymmetric and stationary, and describes the ``exterior field of a charged,
 magnetized, spinning deformed mass'' \cite{MSM}. The MSM spacetime depends on
 five real parameters, namely the mass $m$, the spin (per unit mass)
 $a$, the total charge $q$, the magnetic dipole moment ${\cal M}$, and the
 mass-quadrupole moment ${\cal Q}$. ${\cal M}$ and ${\cal Q}$ are represented in
 the metric by the real parameters  $\mu$ and $b$, i.e.
 \begin{eqnarray} \label{eq:MagQuad}
  {\cal M} &=& \mu+q (a-b)~~, \nonumber \\
  {\cal Q} &=& -m(d-\delta-a~b+a^2)~~, 
 \end{eqnarray}
 where
 \begin{eqnarray} \label{eq:deltad}
   \delta &:=& \frac{\mu^2-m^2 b^2}{m^2-(a-b)^2-q^2}~~, \nonumber \\
   d &:=& \frac{1}{4}[m^2-(a-b)^2-q^2]~~.  
 \end{eqnarray}

 The metric for the MSM spacetime can be given by the Weyl-Papapetrou line
 element 
 \begin{eqnarray}\label{eq:MNSpro}
  ds^2 & = & -f(dt-\omega d\phi)^2 \nonumber \\
       & + & f^{-1} \left[ e^{2\gamma} (d\rho^2+dz^2)+\rho^2 d\phi^2 \right]~~,
 \end{eqnarray}
 where all metric functions $f$,~$\omega$,~and~$\gamma$ are considered as
 functions of the prolate spheroidal coordinates $u,v$, while the coordinates
 $\rho,z$ are the corresponding cylindrical coordinates. The transformation
 between the two coordinate systems is
 \begin{equation} \label{eq:TrCylPS}
  \rho=\kappa \sqrt{(u^2-1)(1-v^2)},~~ z=\kappa u v~~,
 \end{equation}
 where
 \begin{equation} \label{eq:kappa}
  \kappa :=\sqrt{d+\delta}~~. 
 \end{equation}
 
 The metric functions are
 \begin{eqnarray} \label{eq:MetrFunc}
  f &=& E/D~~, \nonumber \\
  e^{2\gamma} &=& E/16 \kappa^8 (u^2-v^2)^4~~,\\
  \omega &=& (v^2-1)F/E~~,\nonumber
 \end{eqnarray}
 where
 \begin{eqnarray} \label{eq:EDF}
  E &=& R^2+\lambda_1\lambda_2 S^2~~, \nonumber \\
  D &=& E+R P+\lambda_2 S T~~, \\
  F &=& R T-\lambda_1 S P~~,\nonumber
 \end{eqnarray}
 \begin{equation} \label{eq:lambda}
  \lambda_1=\kappa^2 (u^2-1),~~\lambda_2=v^2-1
 \end{equation}
 and
 \begin{eqnarray}\label{eq:PRST}
  P & := & 2 \{\kappa m u [(2 \kappa u+m)^2-2 v^2 (2 \delta +a b -b^2)\nonumber \\
    & - & a^2+b^2 -q^2]-2 \kappa^2 q^2 u^2-2 v^2 (4 \delta d-m^2 b^2)\}~~, 
   \nonumber \\
  R & := & 4 [\kappa^2 (u^2-1)+\delta (1-v^2)]^2 \nonumber \\
    &  + & (a-b) [(a-b)(d-\delta)-m^2 b +q~\mu](1-v^2)^2~~,\nonumber \\
    & ~ &  \\
  S & := & -4 {(a-b)[\kappa^2(u^2-v^2)+2 \delta v^2]+v^2 (m^2 b-q~\mu)}~~,
  \nonumber \\
  T & := & 4(2 \kappa m b u+2 m^2 b-q~\mu)[\kappa^2 (u^2-1)+\delta (1-v^2)]
  \nonumber \\
    & + & (1-v^2)\{(a-b)(m^2 b^2 -4 \delta d)  \nonumber \\
    & - & (4 \kappa m u+2 m^2-q^2) [(a-b)(d-\delta)-m^2 b +q~\mu]\}. \nonumber 
 \end{eqnarray}

 The MSM metric is of astrophysical interest as it has been suggested to be 
 appropriate to model neutron stars \cite{MSM,Berti}. By investigating the
 dynamics of the geodesic orbits in the MSM spacetime, Dubeibe et al.
 \cite{Dubeibe07} and Han \cite{Han08} debated on the appearance and on the
 significance of chaos in the MSM model. Even if the main purpose of our article
 is to present a new integration scheme, in section \ref{sec-V} we contribute to
 the aforementioned debate. 

 \subsection{Geodesic motion} \label{subsec-II-b}

 The equations of geodesic motion of a ``test'' particle of rest mass $m_0$ in a
 spacetime given by the metric $g_{\mu\nu}$ are produced by the Lagrangian
 function
 \begin{equation} \label{eq:LagDef}
  L=\frac{1}{2}~m_0~g_{\mu\nu}~ \dot{x}^{\mu} \dot{x}^{\nu}~~,
 \end{equation}
 where the dot denotes derivation with respect to proper time $\tau$. The
 Lagrangian (\ref{eq:LagDef}) has a constant value $L=-m_0/2$ along a
 geodesic orbit, due to the four-velocity
 $g_{\mu\nu}~\dot{x}^{\mu}\dot{x}^{\nu}=-1$ constraint.

 Since the spacetime is axisymmetric and stationary, the corresponding momenta
 \begin{equation}\label{eq:MomDef}
  p_\nu=\pd{L}{\dot{x}^\nu}
 \end{equation}
 are conserved. These are the specific energy
 \begin{equation} \label{eq:EnCon}
  E=-\frac1{m_0}\pd{L}{\dot{t}}~~,
 \end{equation}
 and the specific azimuthal component of the angular momentum
 \begin{equation} \label{eq:AnMomCon}
  L_z =\frac1{m_0}\frac{\partial L}{\partial \dot{\varphi}}~~.
 \end{equation}
 For brevity, we refer hereafter to these two integrals simply as the energy $E$
 and the angular momentum $L_z$. Due to these two integrals of motion, we can
 restrict our study to the meridian plane. We just have to re-express
 \eqref{eq:EnCon}, \eqref{eq:AnMomCon} in order to get $\dot{t}$ and $\dot{\phi}$
 as functions of $E$ and $L_z$, and then replace $\dot{t}$, $\dot{\phi}$ in
 the two remaining equations of motion. Then, from the original set of $4$
 coupled second order ordinary differential equations (ODEs), we arrive to a
 set of $2$ coupled ODEs. 

 By simply applying the Legendre transform
 \begin{equation}\label{eq:LegTr}
  H=p_\mu \dot{x}^\mu-L
 \end{equation}
 to the Lagrangian function (\ref{eq:LagDef}), we get the Hamiltonian function
 \begin{equation}\label{eq:HamDef}
  H=\frac{1}{2m_0}g^{\mu\nu} p_\mu p_\nu~~,
 \end{equation}
 where the momenta $p_\mu$ are given by (\ref{eq:MomDef}) and
 $p^\nu=m_0 \dot{x}^\nu$. The Hamiltonian equations are
  \begin{align}\label{eqn-pdot_qdot}
   \dot{x}^\mu &= g^{\mu\nu}\pd{H}{p^\nu}~~, \nonumber \\
               &~                           \\
   \dot{p}^\mu &=-g^{\mu\nu}\pd{H}{x^\nu}~~. \nonumber
  \end{align}
 As the system is  autonomous
 ($\frac{d H}{d \tau}=\frac{\partial H}{\partial \tau}=0$), the Hamiltonian
 function is an integral of motion and equal to $H=-m_0/2$.
 
\section{Standard symplectic schemes}\label{sec-III}

 We have seen in section \ref{subsec-II-b} that the geodesic equations of motion
 can be described by a Hamiltonian formalism. Thus, symplectic schemes should be
 appropriate for integrating these equations.  

 \subsection{Notation}\label{subsec not}
 In the rest of the paper we drop the Tensor Analysis covariant and
 contravariant symbolism by indices and exponents, and we adopt a symbolism more 
 convenient for Numerical Analysis. By bold characters, we denote vectors, and,
 by arrows over characters, we denote vectors of vectors.
 
  We define
  \begin{align*}
    \mathbf y &:=\begin{pmatrix}\mathbf{p}\\ \mathbf{x}\end{pmatrix}~~, \\    
    J &:=\begin{pmatrix} 0&I\\-I&0\end{pmatrix}~~, \\
    f(\mathbf{y}) &:=J^{-1}\grad{H(\mathbf{y})}~~,
  \end{align*}
   and write the given Hamiltonian system as
  \begin{align}\label{eqn-eom}
    \frac{d \mathbf{y}}{d \tau}=f(\mathbf{y})~~.
  \end{align}

 Further, $\Phi_h$ denotes an integration step of step size $h$. If the
 current position in phase space is $\mathbf{y}_n$, then $\Phi_h$
 propagates the system to the next position $\mathbf{y}_{n+1}$, i.e.
 \begin{align*}
   \mathbf{y}_{n+1}=\Phi_h(\mathbf{y}_n)~~.
 \end{align*}
 To simplify the Runge-Kutta schemes notation we define for an $s$-stage scheme
\begin{itemize}
 \item{}the vector 
\begin{align}
  \vec y_n&:=\underbrace{\begin{pmatrix}\mathbf y_n&...&\mathbf y_n\end{pmatrix}^T}_{\text{s times}}~~,\nonumber
\end{align}
 \item{}the inner-stage values
\begin{align}
  \vec Y &:=\begin{pmatrix} \mathbf{Y}_1&...&\mathbf{Y}_s\end{pmatrix}^T~~,\nonumber
\end{align}
 \item{}the auxiliary variables
\begin{align}\label{not-Z_F_A}
  \vec Z &:=\begin{pmatrix} \mathbf{Z}_1&...&\mathbf{Z}_s\end{pmatrix}^T=\vec Y-\vec y_n~~,
\end{align}
  \item{}the function
\begin{align}
  F(\vec Y) &:=  \begin{pmatrix} f(\mathbf{Y}_1)&... &
  f(\mathbf{Y}_s)\end{pmatrix}^T~~,\nonumber
\end{align}
  \item{}and the coefficient matrix
\begin{align}
  A &:=\begin{pmatrix}a_{11}&\hdots&a_{1s}\\ \vdots&\ddots&\vdots
   \\ a_{s1}&\hdots&\ a_{ss}\end{pmatrix}~~.\nonumber
\end{align}
\end{itemize}

 The exclamation mark ``!'' over relation symbols denotes requirement.
 Furthermore, $\cal X$ denotes the phase space and $\vec{\cal X}$ denotes the
 space on which the $\vec Z$ are defined, i.e.
 $\vec{\cal X}:=\cal{X} \times \hdots \times \cal{X}$.

 As a norm for matrices we use the \textit{Frobenius norm}, which for a matrix
 $A$ is
  \begin{align*}
   \|A\|_{\text{Frob}}=\sqrt{\sum_{i,j}A_{ij}^2}~~.
  \end{align*}

 \subsection{The equations of motion from a numerical point of view}

 From a numerical point of view, it is easier to handle the Hamiltonian equations
 $\eqref{eqn-pdot_qdot}$ in the form $\eqref{eqn-eom}$. The Hamiltonian system
 $\eqref{eq:HamDef}$ is non-separable in the sense that the mass-matrix in the
 Hamiltonian depends on the positions $\mathbf{x}$ and, therefore, the momenta
 cannot be separated from the position coordinates. In the case of a
 non-integrable system, the set of coupled first order differential equations
 $\eqref{eqn-eom}$ can be very sensitive to changes in the arguments
 $\mathbf{y}$. This happens when we evolve  chaotic orbits by $\Phi_h$. In
 other words, if $Df(\mathbf{y})$ denotes the Jacobian of $f(\mathbf{y})$ at
 a point $\mathbf{y}$ of the phase-space $\cal X$, then there
 exists a subset ${\cal U} \subset {\cal X}$, so that
 \begin{align}\label{eqn-inequal_DF}
  \|Df(\mathbf{y})\|\gg 1 ~~~\text{$\forall~\mathbf{y}\in {\cal U}$}~~.
 \end{align}

 As the Hamiltonian system $\eqref{eq:HamDef}$ is not separable, a natural
 choice for a symmetric, symplectic integrator is a \textit{Gauss-collocation
 scheme} (see e.g., \cite{hairernorsettwanner}).
 \subsubsection{Collocation methods}

 Given an interval $[0,h]$, stages $0\le c_1<...<c_s\le 1$ and an initial-value
 problem
 \begin{align}
    &y(0) = y_0\nonumber\\
    &\dot{y} = f(\tau,y)~~,\nonumber
 \end{align}
 the polynomial $U(\tau)$ of degree $s$, satisfying
  \begin{align}
     &U(0) =y_0\nonumber\\
     &\dot{U} (c_ih) =f(c_ih,U(c_ih)),~~i=1,...,s~~,\nonumber
  \end{align}
 is called a \textit{collocation polynomial}. A \textit{collocation method}
 consists of finding such a $U(\tau)$ and then setting
  \begin{align}
    y(h)\approx U(h).
  \end{align}
 It can easily be shown (e.g., theorem 7.7. in \cite{hairernorsettwanner}) that
 a collocation method as described above is equivalent to an $s$-stage
 Runge-Kutta scheme  
 \begin{align}\label{eqn-Zi}
  \mathbf{y}_{n+1} &= \mathbf{y}_n+h\sum_{i=1}^sb_if(\mathbf{Y}_i)~~,
  \nonumber\\
  \mathbf{Y}_i &=\mathbf y_n+ h\sum_{j=1}^sa_{ij}f(\mathbf{Y}_j)~~,
 \end{align}
 with coefficients
 \begin{align}
    a_{ij}=\int_0^{c_i}l_j(\tau)d\tau\nonumber\\
    b_i=\int_0^1l_i(\tau)d\tau\nonumber,
  \end{align}
 where $l_i(\tau)$ are the Lagrange-polynomials
 \begin{align*}
  l_i(\tau)=\frac{\prod_{i\neq j}(\tau-c_j)}{c_i-c_j}~~.
 \end{align*}
 If the inner stages $c_i$ are chosen as 
 \begin{align*}
	c_i=\frac 1 2(1+\tilde c_i)~~,
 \end{align*}
 where $\tilde c_i$ are the roots of the Legendre-polynomial of degree $s$, one
 obtains a Gauss-collocation scheme, which can be shown to be of the highest
 possible order $\mathcal O(h^{2s})$. Furthermore, Gauss-collocation or
 \textit{Gauss-Runge-Kutta methods} are symmetric and symplectic, see
 \cite{hairerlubichwanner}, chapter V and VI.

 If we rewrite the system of implicit eqs. $\eqref{eqn-Zi}$ in matrix-vector
 notation, we get
 \begin{align}
  \begin{pmatrix}\mathbf Y_1\\ \vdots\\ \mathbf Y_s\end{pmatrix}=\begin{pmatrix}\mathbf y_n\\
  \vdots\\ \mathbf y_n\end{pmatrix}+h\begin{pmatrix}a_{11}I&...&a_{1s}I\\
  \vdots&\ddots&\vdots\\ a_{s1}I&...&a_{ss}I\end{pmatrix}
 \begin{pmatrix}f(\mathbf Y_1)\\ \vdots \\f(\mathbf Y_s)\end{pmatrix}\label{eqn-matrix-vector}~~,
 \end{align}
 with the $s\times s$-identity matrix $I$. With the help of the direct product
 of matrices $\otimes$ and the notations in $\ref{subsec not}$, 
 eq. $\eqref{eqn-matrix-vector}$ becomes 
 \begin{align}\label{eqn-Y}
  \vec Y=\vec y_n+h(A\otimes I)F(\vec Y)~~.
 \end{align}
 Following \cite{hairerwanner}, chapter IV.8, we use the auxiliary variables
 defined in eqs. $\eqref{not-Z_F_A}$, to get the system in a shorter form 
 \begin{align}\label{eqn-Z}
  \vec Z=h(A\otimes I)F(\vec Z)~~,
 \end{align}
 where $F(\vec Z)$ is to be read as  
 \begin{align}\label{eqn-FZ}
  F(\vec Z)=\begin{pmatrix}f(\mathbf Y_1)\\ \vdots \\f(\mathbf Y_s)\end{pmatrix}
  =\begin{pmatrix}f(\mathbf y_n+\mathbf Z_1)\\
   \vdots \\f(\mathbf y_n+\mathbf Z_s)\end{pmatrix}~~.
 \end{align}
 There are two possibilities to solve this implicit equation for the inner stage
 values $\vec Z$.
 \begin{itemize}
 \item The simplest and the most popular method is the
 \textit{Fixed-Point iteration}. One has to solve iteratively
  \begin{align}
   \vec Z^{k+1}=h(A\otimes I)F(\vec Z^k)~~.
  \end{align}
  The \textit{Banach fixed-point theorem} guarantees that the iteration converges
  towards the correct $\vec Z$, if it is a contraction. This requires
  \begin{align*}
    \|\vec Z^{k+2}-\vec Z^{k+1}\| \stackrel{!}{<}\|\vec Z^{k+1}-\vec Z^k\|~~.
  \end{align*}
  Because of
  \begin{align*}
   & \|\vec Z^{k+2}-\vec Z^{k+1}\|=\|h(A\otimes I)(F(\vec Z^{k+1})-F(\vec Z^k))\|\\
   & \le \max_{\vec{Z}\in \vec{\cal X}}~
   \|h(A\otimes I)DF(\vec Z)\|\|\vec Z^{k+1}-\vec Z^k\|~~,
   \end{align*}
  this implies
  \begin{align} \label{eqn-req_fpi}
   h\|(A\otimes I)DF(\vec{Z})\|\stackrel{!}{<}1,
     ~~~\forall~\vec{Z}\in \vec{\cal X}~~.
  \end{align}
 \item If the convergence of the fixed-point iteration is not ensured, we can
 look forward to get better accuracy by searching for the roots of
 \begin{align}
  \hat F(\vec Z):=\vec Z-h(A\otimes I)F(\vec Z)
 \end{align}
 via a \textit{modified Newton iteration}. In this case, we have to iterate
 \begin{align}
  \vec Z^{k+1} &= \vec Z^{k}+\Delta \vec Z^{k}\\
  \Delta\vec Z^k     &= -M^{-1}\hat F(\vec Z^k)~~\label{eqn-DeltaZ},
 \end{align}
 with
 \begin{align}
  M:=I-(A\otimes I)( I\otimes Df(\mathbf y_n))~~.
 \end{align}
 Again, convergence can only be ensured for a contraction. Therefore, we need
 \begin{align*}
  &\|\vec Z^{k+2}-\vec Z^{k+1}\|= \\
  &\|\vec Z^{k+1}-M^{-1}\hat F(\vec Z^{k+1})-\left(\vec Z^k-M^{-1}\hat
  F(\vec Z^k)\right)\|    \\ 
  & \le \max_{\vec{Z}\in \vec{\cal X}}~h\|I-M^{-1}D\hat F(\vec{Z})
  \|\|\vec Z^{k+1}-\vec Z^k\| \\
  & \stackrel{!}{<} \|\vec Z^{k+1}-\vec Z^k\|~~,
 \end{align*}
 or
 \begin{align} \label{eqn-req_ni}
  h\|I-M^{-1}D\hat F(\vec{Z})\|\stackrel{!}{<}1,
  ~~\forall ~\vec{Z}~\in~ \vec{\cal X}~~.
 \end{align}
 \end{itemize}
 As a single iteration step for the fixed-point iteration is computationally
 much cheaper than in the modified newton iteration case, one should prefer the
 first whenever it is applicable. In both cases, inequality
 $\eqref{eqn-inequal_DF}$ together with the respective requirements
 $\eqref{eqn-req_fpi}$ or $\eqref{eqn-req_ni}$ lead to severe restrictions on
 the possible step-size $h$. The \textit{ultima ratio} for the solution of
 eq. $\eqref{eqn-Z}$ is the use of a \textit{Newton-Raphson} method. There, $M$
 in eq. $\eqref{eqn-DeltaZ}$ is replaced by 
 \begin{align*}
	M(\vec Z^k)=I-(A\otimes I)(DF(\vec Z^k))~~.
 \end{align*}
 This gives an iteration with quadratic convergence but is much more expensive
 because for every iteration, a Jacobian has to be calculated. Therefore,
 standard literature recommends to proceed without it whenever it is possible, e.g.
 \cite{hairerlubichwanner}, chapter VIII. We will see in section $\ref{sec-V}$,
 that the use of Newton-Raphson doesn't improve the performance of the integration
 scheme when it is applied in the case of geodesic orbits either.

 \subsection{Existing step-size controls} \label{subsec-existing}

 For efficiency's sake one would expect to use small time steps only for the
 subspace for which condition $\eqref{eqn-inequal_DF}$ holds. Therefore, a
 reliable step size control is necessary. A feasible step size  control has to
 preserve the geometric properties of the underlying integration scheme. As it
 was shown in \cite{stoffer88}, there cannot be an efficient symplectic
 integrator with adaptive step size. Thus, other structure preserving properties
 of the integrator, such us \textit{symmetry}, i.e.
 \begin{align}\label{eqn-symmetry1}
  \Phi_{-h}(y)=\Phi^{-1}(y)
 \end{align}
 and \textit{reversibility}, i.e.
 \begin{align}\label{eqn-reversibility1}
  \Phi^{-1}_h\circ \zeta =\zeta \circ \Phi_h 
 \end{align}
 for the involution $\zeta(\mathbf{p},\mathbf{x})=(-\mathbf{p},\mathbf{x})$,
 have to be conserved. If we apply such an integrator to a symmetric, reversible
 differential equation, we would expect the same advantageous long term behavior
 as for a symplectic integrator, e.g., \cite{hairerlubichwanner}, chapter VIII.

 One of the most efficient step size controls has been proposed by Hairer and
 S\"oderlind \cite{soederlind}. They express the discrete variable time step 
 sequence $\tau={\tau_1,...,\tau_n,\tau_{n+1},...}$ by means of a constant step
 size sequence $\epsilon={\epsilon_1,...,\epsilon_n,\epsilon_{n+1},...}$ and a
 scalar function $\sigma(\mathbf y)$ via $d\tau=\sigma(\mathbf y)d\epsilon$. For
 the dependence of $\mathbf{y}$ on $\epsilon$, they obtain

 \begin{align}\label{eqn-dy_dtau}
  \frac{d \mathbf{y}}{d \epsilon}=\frac{d \mathbf{y}}{d \tau} \d \tau \epsilon
  =f(\mathbf{y})\d \tau \epsilon~~.
 \end{align}

 Defining
 \begin{align}\label{eqn-w_hairer}
   w:=\frac 1{\sigma(\mathbf{y})}~~,
 \end{align}
 one finds
 \begin{align}\label{eqn-w_dot_hairer}
   \frac{d w}{d\epsilon}&=-\frac1{\sigma^2}\grad\sigma \frac{d \mathbf{y}}{d \epsilon} \\
     &=-\frac1{\sigma}\grad\sigma f(\mathbf{y})~~.
 \end{align}
 We now combine eqs. $\eqref{eqn-dy_dtau}$ and $\eqref{eqn-w_dot_hairer}$ and
 get the system
 \begin{align}\label{eqn-augmented_hairer}
  \begin{pmatrix}
   \frac{d\mathbf{y}}{d \epsilon}\\
   \frac{d w}{d \epsilon}
  \end{pmatrix}=
  \begin{pmatrix}
   \frac 1 w f(\mathbf{y})\\
   -\frac1{\sigma}\grad\sigma f(\mathbf{y})=:G(\mathbf{y})
  \end{pmatrix}~~.
 \end{align}

 Solving this system with the scheme
 \begin{align}
  w_{\text{$n+\frac 1 2$}} &=w_n+\epsilon \cdot G(\mathbf{y}_n) \label{eqn-w_hairer_num}\\
  \mathbf{y}_{\text{$n+1$}} &= \Phi_{\frac{\epsilon}{w_{n+\frac 1 2}}}(\mathbf{y}_n)\\
  w_{\text{$n+1$}} &=w_{\text{$n+\frac 1 2$}}+\epsilon \cdot G(\mathbf{y}_{n+1})~~,\\
  w_0 &=\frac1{\sigma(\mathbf{y}_0)}
 \end{align}
 we get a symmetric, reversible integrator, as it is shown in \cite{soederlind}.
 Nevertheless, this step-size algorithm has some drawbacks which can show up when 
 integrating geodesic equations of motion of a non-integrable system.

 The above integration scheme is explicit in the variable $w$. Therefore, eq.
 $\eqref{eqn-w_hairer_num}$ represents a constraint on possible step size
 adapters $\sigma(\mathbf{y})$. If we don't want the underlying step size
 $\epsilon$ to be too small, $G(\sigma(\mathbf{y}))$ has to be bounded. This is
 similar to the case when we use explicit Runge-Kutta schemes for stiff
 differential equations, where, due to a very small range of stability, the
 step size has to be unfeasibly small.

 However, for the iterations to converge, we have to choose a step size small
 enough to get a contraction.
 \begin{itemize}

  \item If we want to solve equation $\eqref{eqn-Z}$ with a
  fixed-point-Iteration, the step size must be somehow proportional to
  the inverse of $\|Df(\mathbf{y})\|$, i.e. we have
  \begin{align}
   \sigma(\mathbf{y})\propto\frac1{\|Df(\mathbf{y})\|}~~,
  \end{align}
  and such a controller $\sigma$ yields
 \begin{align}
  G(\mathbf{y})=\frac{\sum_{ijk}\left(Df(\mathbf{y})\right)_{ij}
  \left(\pd{\left(Df(\mathbf{y})_{ij}\right)}
  {\mathbf{y}_k}\right)\left(f(\mathbf{y})\right)_k}
  {\|Df(\mathbf{y})\|^2}~~.
 \end{align}

 For regions of the phase space, where
 \begin{align}
  \|D^2f\cdot f\|\gg\|Df\|~~,
 \end{align}
 this would conflict with the requirement which demands $G(\mathbf{y})$ not to
 be too large. Numerical tests even show that in such cases the step size can
 become negative.

 \item In order to guarantee the convergence of the newton-iteration, i.e. to
 satisfy condition $\eqref{eqn-req_ni}$, it is required that
 \begin{align}
  & \sigma(\mathbf{y}) \propto  \nonumber \\
  & \frac1{\|I-M^{-1}D\hat F(\vec{Z})\|} = \nonumber \\
  & \frac1{\|I-(I-(A\otimes Df(\mathbf{y}_n)))^{-1}(I-(A\otimes I)DF(\vec{Z}))\|}~~. 
  \nonumber \\
   & ~ 
 \end{align}
 It is by no means clear how one can consider the last expression as a function
 of $\mathbf y$. The only possibility would be to set
 $\vec Z=(\mathbf 0\hdots\mathbf 0)^T$ in the last expression. The calculation
 of the corresponding $G(\mathbf{y})$ would then lead to an expression with a 
 factor
 \begin{align}
  \frac1{\|I-(I-(A\otimes Df(\mathbf{y}_n)))^{-1}(I-(A\otimes I)DF(\vec Z))\|^2}~~.\nonumber
 \end{align}  
 Because of $DF(\vec Z)=I\otimes Df(y)$ for the $\vec Z$ chosen above, this
 factor would be
 \begin{align}\label{eqn-factor_w_hairer}
  \frac1{\|I-(I-(A\otimes Df(\mathbf{y}_n)))^{-1}(I-(A\otimes Df(\mathbf y)))\|^2}~~.
 \end{align}  
 Thus, the evaluation of the function $G$ at the point $\mathbf{y}_n$ (which is
 necessary in eq. $\eqref{eqn-w_hairer_num}$) is impossible as we would have to
 divide by zero due to the factor $\eqref{eqn-factor_w_hairer}$. We see that
 it is not possible to use the modified newton iteration along with the 
 presented step size control algorithm.

\end{itemize} 

 \section{A new adaptive step size control of motion}
 \label{sec-IV}
 
 We now present a new integration scheme $\Phi_h$ with an adaptive step size
 $h(\epsilon,\mathbf{y})$. The integrator uses an $s$-stage Gauss-collocation
 method as the underlying integrator. The step size depends on an underlying
 constant step size $\epsilon$ and the actual state of the system $\mathbf{y}$. 
 The integration scheme has the following properties:

 \begin{itemize}

 \item For the step size $h$ an equation similar to
  \begin{align}
   h(\epsilon,\mathbf{y})\propto \frac1{\|Df(\mathbf{y})\|}
  \end{align}
 holds, and, thus, condition $\eqref{eqn-req_fpi}$ is satisfied. 
  
 \item The integrator is both symmetric and reversible. Hence, the theoretical
 results on the long term behavior of such integrators are applicable
 (e.g., \cite{hairerlubichwanner}, chapter XI). 

 \end{itemize}
  
 The main idea of the step size control algorithm is to replace requirement
 $\eqref{eqn-req_fpi}$ by
 \begin{align}
  \|h\cdot Df(\mathbf{y})\|\stackrel !=\epsilon~~,
 \end{align}
 for some $\epsilon<1$. We slightly modify this formulation to
 \begin{align}\label{eqn-h}
  \|\frac h 2\left(Df(\mathbf{Y}_1)
  +Df(\mathbf{Y}_s)\right)\|\stackrel !=\epsilon~~,
 \end{align}
 with $\mathbf{Y}_{1(s)}$ being the inner-stage value at stage $1(s)$ of the
 Gauss-Runge-Kutta scheme. In the subsequent paragraphs we demonstrate that this
 modification gives the required structure preserving properties.

 For the integrator to be symmetric, we must make sure that when we propagate the
 system backwards in time (i.e. from $\mathbf{y}_{n+1}$ to $\mathbf{y}_n$) the step
 size has the same behavior as when propagating forward (i.e. from  $\mathbf{y}_n$
 to $\mathbf{y}_{n+1}$), i.e. we need
 \begin{align}
  h(\epsilon,\mathbf{y}_n)=-h(-\epsilon,\mathbf{y}_{n+1})~~.
 \end{align}

 This, together with the symmetry of the underlying integrator (i.e. the
 validity of property $\eqref{eqn-symmetry1}$ for constant steps $h$), ensures
 the symmetry of the whole integration scheme. Let $ \mathbf{\hat Y}_i$ denote
 the inner-stage values for the integration backwards in time. The symmetry of
 the Gauss-collocation method then results in
 \begin{align}\label{eqn-symmetry_Z}
  \mathbf{\hat Y}_i=\mathbf{Y}_{s+1-i}
  ~~\forall i=1,\hdots,s~~,
 \end{align}
 because the collocation polynomial for $\mathbf{y}_{n+1}=\Phi_h(\mathbf{y}_n)$,
 which is the interpolation polynomial through the points $(0,y_n)$ and 
 $(c_i,\mathbf{Y}_{i})$, is the same as the
 collocation polynomial for $\mathbf{y}_{n}=\Phi_{-h}(\mathbf{y}_{n+1})$, i.e.
 the interpolation polynomial through the points $(1,y_{n+1})$ and  
 $(c_i,\mathbf{\hat Y}_{i})$. Due to property
 $\eqref{eqn-symmetry_Z}$, we get
 \begin{align}
  h(-\epsilon,\mathbf{y}_{n+1}) 
  &= \frac{-\epsilon}{\|\frac 1 2[Df(\mathbf{\hat Y}_1)+Df(\mathbf{\hat Y}_s)]\|} \nonumber \\
  &= \frac{-\epsilon}{\|\frac 1 2[Df(\mathbf{Y}_s)+Df(\mathbf{Y}_1)]\|} \nonumber \\
  &= -h(\epsilon,\mathbf{y}_n)~~.
 \end{align}

 If we consider the new variable step size integrator
 $\Phi_h(\epsilon,\mathbf{y})$ as a constant step size integration scheme
 $\Psi_{\epsilon}$, then the reversibility condition $\eqref{eqn-reversibility1}$
 reads
 \begin{align}\label{eqn-reversibility2}
  \Psi^{-1}_\epsilon\circ \zeta =\zeta \circ \Psi_\epsilon~~.
 \end{align}
 
 If we have
 \begin{align}\label{eqn-h_reversible}
  h(-\epsilon,\zeta~\mathbf{y}_{n+1})=-h(\epsilon,\mathbf{y}_n)~~,
 \end{align}
 then the reversibility of the method is a direct consequence of the
 reversibility of the underlying integrator $\eqref{eqn-reversibility1}$. In
 order to prove condition $\eqref{eqn-h_reversible}$, we denote by
 $\mathbf{\hat Y}$ the inner stage values for the integration that starts at
 $\zeta~\mathbf{y}_{n+1}$. We then notice, that the reversibility of the
 Gauss-Runge-Kutta scheme implies 
 \begin{align}
  \mathbf{\hat Y}_i=\zeta\mathbf{Y}_{s+1-i}~~.
 \end{align}
 This yields
 \begin{align}
  h(-\epsilon,\zeta \mathbf{y}_{n+1}) &=
  \frac{-\epsilon}{\|\frac 1 2[Df(\mathbf{\hat Y}_1)+
   Df(\mathbf{\hat Y}_s)]\|} \nonumber \\
  & =\frac{-\epsilon}{\|\frac 1 2\zeta^{-1} [Df(\mathbf{Y}_s)
    +Df(\mathbf{Y}_1)]\|} \nonumber \\
  & =-h(\epsilon,\mathbf{y}_n)~~,
 \end{align}
 where the last equality is a consequence of the orthogonality of the involution
 $\zeta$.

 Equation $\eqref{eqn-h}$, together with equation $\eqref{eqn-Z}$ for the
 inner-stage values $\mathbf{Z}_i$, leads to the system of equations
 \begin{align}\label{eqn-impl-IGEM}
  \begin{pmatrix} \vec Z\\h\end{pmatrix}=\begin{pmatrix}h(A\otimes I)F(\vec Z)
  \\\frac{\epsilon}{\|\frac 1 2[Df(\mathbf y_n+\mathbf{Z}_1)+Df(\mathbf y_n+\mathbf{Z}_s)]\|}\end{pmatrix}~~.
 \end{align}

 Inserting the second part into the first, we can easily apply a
 fixed-point-Iteration, which yields
 \begin{align} \label{eqn-FP-IGEM}
  \vec Z^{k+1}=\frac{\epsilon(A\otimes I)F(\vec Z^k)}
     {\frac 1 2\|Df(\mathbf y_n+\mathbf{Z}_1^k)+Df(\mathbf y_n+\mathbf{Z}_s^k)\|}~~. 
 \end{align}

  \begin{figure*} [htp]
   \centering
   \includegraphics[width=0.8\textwidth]{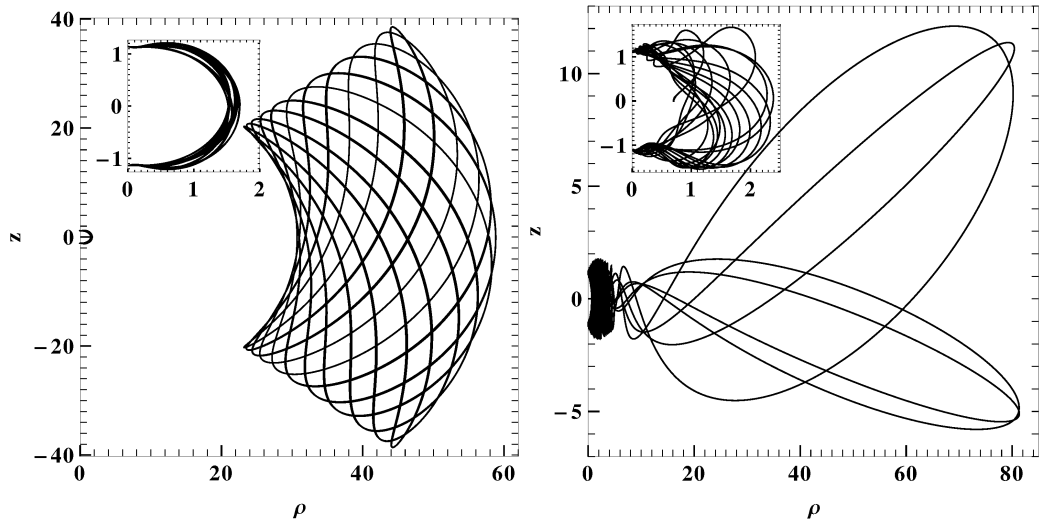}
    \caption{For $m=2.904$, $a=1.549$, $q=0$, $\mu=0$, $b=0.8$, $E=0.971$, and
   $L_z=9.3$, the left panel shows the projections on the $(\rho,z)$ plane of
   two regular orbits, and the right one projections of a chaotic orbit. In the
   left panel the initial conditions of the left orbit (magnified at the upper left
   embedded plot) are $\rho=1.7$, $z=\dot{\rho}=0$, while the initial conditions
   for the right regular orbit are $\rho=30.7$, $z=\dot{\rho}=0$. In the right
   panel, the initial conditions for the chaotic orbit are $\rho=0.7$,
   $z=\dot{\rho}=0$. A magnification of the left part of the chaotic orbit is shown
   in the embedded upper left plot.}
   \label{FigOr}
  \end{figure*} 

 \section{Numerical experiments}\label{sec-V}

 In this section we test our Integrator for Geodesic Equations of Motion (IGEM)
 and compare it with several other integration schemes. These are

 \begin{itemize}

  \item A standard fifth order \textit{Cash-Karp-Runge-Kutta} scheme with
   constant step size $\epsilon$ (RK5con) as proposed by Numerical Recipes
   \cite{nr}.

  \item The Cash-Karp-Runge-Kutta scheme (RK5var) with initial step size
  $\epsilon$ and a step size algorithm as follows:

   \begin{itemize} 
   
    \item Starting with $\mathbf{y}_n$ calculate
            $\mathbf{y}_{n+1}=\Phi_h(\mathbf{y}_n)$

    \item Calculate the Hamiltonian's relative error 
  $\delta H=\abs{\frac{H(\mathbf{y}_{n+1})-H(\mathbf{y}_{n})}{H(\mathbf{y}_{n})}}$, 
   between two consecutive integration points $\mathbf{y}_{n},~\mathbf{y}_{n+1}$.

    \item 
     \begin{itemize}
      \item If $\delta H$ is greater than a minimum tolerance value
      $tol_1=10^{-12}$, then repeat the integration step with $h_{new}=\frac h 2$.

      \item If $\delta H$ is smaller than a second tolerance value
      $tol_2=10^{-14}$, then double the step size $h$ before calculating the
      next step.

      \item If $tol_2\le\delta H\le tol_1$, then calculate the next step.
     \end{itemize}
   \end{itemize} 

  \item A standard symplectic integration scheme used in classical celestial
  mechanics given by the Gauss-collocation as underlying integrator together
  with the step size control explained in section $\ref{subsec-existing}$ with
  $\sigma(\mathbf{y})=\frac1{\|f(\mathbf{y})\|}$ (CCM). 

  \item Again the standard scheme from classical celestial mechanics with
  $\sigma(\mathbf{y})$ replaced by
  $\sigma(\mathbf{y})=\frac1{\|Df(\mathbf{y})\|}$ (CCM2).

  \end{itemize}

 In all implicit schemes, we use a fixed-point iteration which we implement as
 proposed by \cite{hairer2}:
 In order to increase the accuracy, one iterates until either the difference 
 between two subsequent $\vec Z$ is numerically $0$ or fluctuations due to
 round-off errors start to occur. In other words, we use the stopping criterion 
 \begin{align}
  \|\vec Z^{k+1}-\vec Z^k\|=0~\text{or}~\|\vec Z^{k+1}-\vec Z^k\|> \|\vec Z^k-\vec Z^{k-1}\|
  \nonumber
 \end{align}

 As test case for the algorithms, we take the system described in section
 $\ref{sec-II}$. In particular, we use the set of MSM parameters and initial
 condition discussed in \cite{Han08}. Accordingly, we choose a central object
 with mass $m=2.904$, spin $a=1.549$, and charge $q=0$, while the
 real parameters $\mu$ and $b$ correlated with the magnetic dipole and the
 mass-quadrapole deformation are set to be $0$ and $0.8$ respectively.
 Regarding the initial conditions, the energy and the angular momentum are set
 to be $E=0.971$ and $L_z=9.3$, respectively, while we fix $\dot{\rho}=z=0$.
 The integrators are tested for 3 different values of $\rho$, which correspond
 to 3 different cases of orbits: two regular ones (left panel of Fig.
 \ref{FigOr}), each covering quasiperiodically a corresponding torus in the
 phase space, and one chaotic (right panel of Fig. \ref{FigOr}), which covers in
 an irregular manner all the available phase space. We propagate the
 system in time until the proper time reaches $\tau_f=500000$ and compare the
 cpu times $T_{\text{calc}}$  the calculation lasted. As a measure of our code's
 accuracy, we track the overall relative error
 $\Delta H=\abs{\frac{H(\mathbf{y}_{n})-H}{H}}$, where $H$ is the theoretical
 value, while $H(\mathbf{y}_{n})$ is the computed Hamiltonian value at the
 integration point $\mathbf{y}_{n}$. We abort the simulations if
 $\Delta H>10^{-6}$.

  \begin{figure*} [htp]
   \centering
   \includegraphics[width=0.8\textwidth]{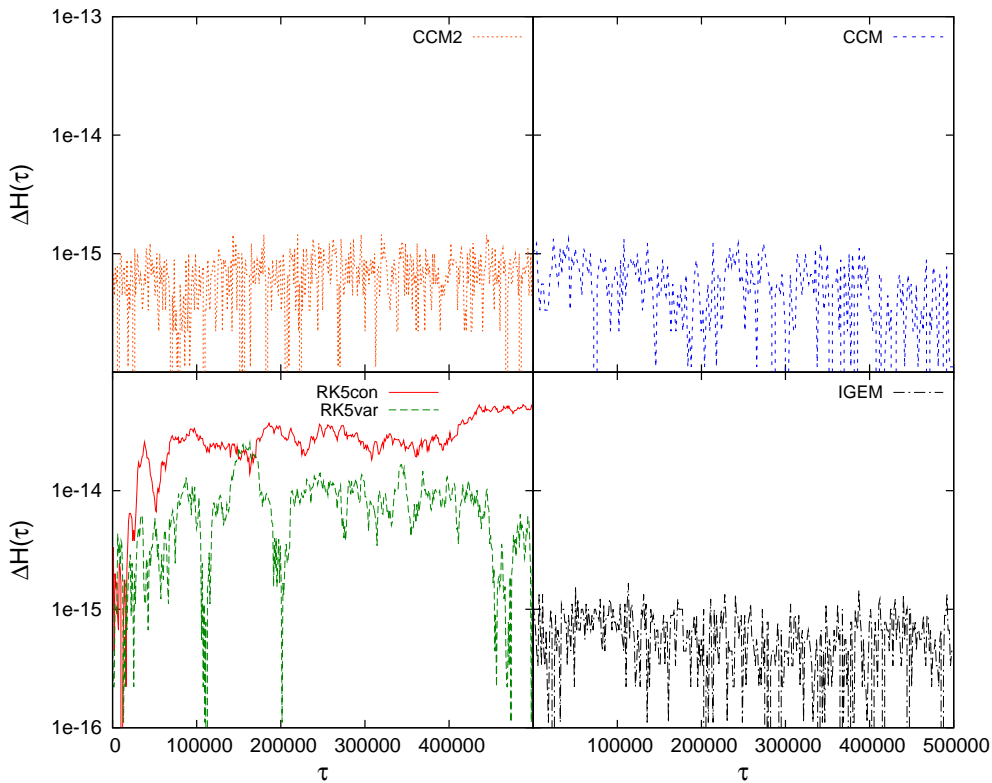}
    \caption{The relative error $\Delta H$ as function of the proper time $\tau$
    in semi-logarithmic scale for the regular orbit with $\rho=30.7$.}
   \label{fig-307-energy}
  \end{figure*} 

 \subsection{Case with $\rho=30.7$}

 The initial condition $\rho=30.7$ corresponds to a regular orbit (right orbit in
 the left panel of Fig. \ref{FigOr}). Fig. $\ref{fig-307-energy}$ shows that the
 error is much larger for the explicit schemes, although we have chosen a much
 smaller step size $\epsilon$. In table $\ref{tab-307}$ we see that not only are
 the explicit integrators less accurate, but they also require much more
 computation time. Moreover, if we increase the step sizes $\epsilon$ of the
 explicit schemes, the energy drift becomes much worse. On the other hand, CCM
 is the fastest algorithm in this case because the corresponding regular orbit does
 not posses the special feature discussed in section \ref{subsec-existing} and
 expressed by condition $\eqref{eqn-inequal_DF}$. Overall, the explicit
 integrators show a linear drift whereas the symmetric integrators preserve the
 Hamiltonian's constant value.
 
  \begin{table}[htp]
   \centering
%  \small
   \begin{tabular}{ccc}
    \toprule
     Integrator & $\epsilon$ & $T_{\text{calc}}[s]$\\
    \midrule
     RK5con & $0.01$ & $222.9$\\
     RK5var & $0.01$ & $160.6$\\
     CCM    & $1.0$ & $17.3$\\
     CCM2   & $1.0$ & $46.1$\\
     IGEM   & $1.0$ & $41.3$\\
    \bottomrule
  \end{tabular}

 \normalsize

  \caption{The cpu calculation times for the proper time interval
 $\tau \in [0,500000]$ for different integration schemes in the case of the 
 regular orbit with $\rho=30.7$.
  }
  \label{tab-307}
 \end{table} 

 \subsection{Case with $\rho=1.7$}

\begin{figure} [htp]
  \centering
  \includegraphics[width=0.4\textwidth]{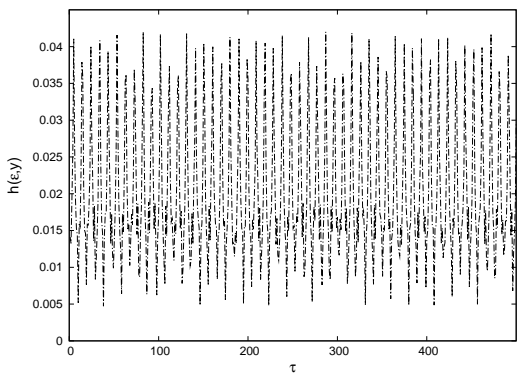}
  \caption{ The step size $h$ of the IGEM as function of proper time
  $\tau\in[0,500]$ for the regular orbit with $\rho=1.7$.}
  \label{fig-17-h(IGEM)}
 \end{figure}

 Even though the initial condition $\rho=1.7$ corresponds to a regular orbit
 (embedded plot of the left panel of Fig. \ref{FigOr}), the system of
 differential equations behaves in a more ill-mannered way than in the
 $\rho=30.7$ regular case. This is shown by tracking the step sizes of the
 IGEM-integrator along the propagation (Fig. $\ref{fig-17-h(IGEM)}$), as we have
 $h\propto\frac 1{\|Df(y(\tau))\|}$. $\|Df(y(\tau))\|$ is varying fast with time.

 \begin{figure} [htp]
  \centering
  \includegraphics[width=0.4\textwidth]{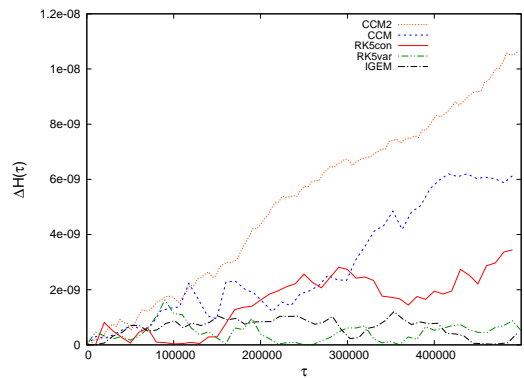}
  \caption{The relative error $\Delta H$ as function of the proper time $\tau$
   in normal scale for the regular orbit with $\rho=1.7$.}
  \label{fig-17-energy}
 \end{figure}  
 \begin{table}[htp]
  \centering
  %\small
 \begin{tabular}{ccc}
  \toprule
   Integrator & $\epsilon$ & $T_{\text{calc}}[s]$\\
  \midrule
   RK5con & $10^{-4}$ & $23813.2$\\
%   RK5var & $10^{-4}$ & $\rightarrow \infty$\\
%   RK5var$_{\text{new $tol$}}$ & $10^{-4}$ & $19207.7$\\
   RK5var & $10^{-4}$ & $19207.7$\\
   CCM & $0.1$ & $266.9$ \\
   CCM2 & $0.1$ & $2100.9$ \\
   IGEM & $0.1$ & $2202.1$\\
  \bottomrule
 \end{tabular}

 \normalsize
 \caption{The calculation times $T_{\text{calc}}$ for the proper time interval
  $\tau \in[0,500000]$ for the different integrators in the case of the
 regular orbit with $\rho=1.7$.}
  \label{tab-17}
 \end{table} 

 When we tested the RK5var scheme, it needed $1$ week of calculation time only
 to arrive at $\tau=758$. Therefore, we relaxed the restrictions on the
 acceptable relative errors per step and set $tol_1=10^{-9}$ and
 $tol_2=10^{-11}$. Now, if we compare the five integrations schemes in table
 $\ref{tab-17}$ and Fig. $\ref{fig-17-energy}$ we notice that the symplectic
 standard schemes from classical celestial mechanics -although fast- show very
 bad conservation properties for the constant of motion. The relative error is
 even worse than for the constant step size RK5con which shows a linear drift. 
 On the other hand, IGEM shows the best long term behavior for affordable
 computational cost. 

 \subsection{Case with $\rho=0.7$}

 \begin{figure} [htp]
  \centering
  \includegraphics[width=0.4\textwidth]{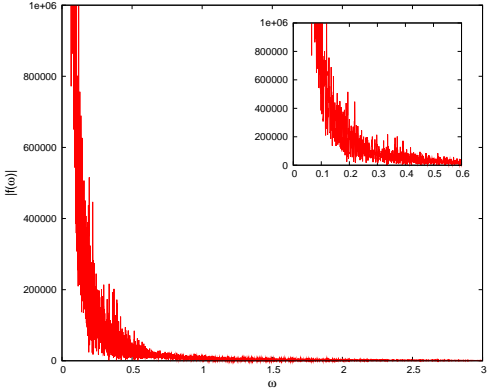}
  \caption{The spectrum of frequencies
  $\abs{f(\omega)}=\abs{\int{e^{i\omega\tau}}\rho(\tau)d\tau}$,
  shows the characteristic noise of chaotic orbits. A magnification of the Fourier
 spectrum is shown in the embedded plot.}
  \label{fig-07-spec}
 \end{figure} 

 \begin{table*}[htp]
  \centering
  %\small
  \begin{tabular}{cccc}
  \toprule
   Integrator & $\epsilon$ & result & $T_{\text{calc}}[s]$ until abortion\\
  \midrule
   RK5con & $10^{-6}$ & aborted after propagation time of $\tau=11254.0$ because
   $\Delta H>10^{-6}$ & $40739.7$ \\
   RK5var & $10^{-4}$ & aborted after propagation time of $\tau=1087.3$ because
   $\Delta H>10^{-6}$ & $149469.9$\\
   CCM & $0.01$ & aborted after propagation time of $\tau=9559.9$ because
   $h(l,\mathbf{y})<0$ & $34.2$\\
   CCM2 & $0.01$ & aborted after propagation time of $\tau=4180.1$ because
   $\Delta H>10^{-6}$ & $1353.9$\\
   IGEM & $0.1$ & no abortion in $\tau=[0,50000]$, $T_{\text{calc}}=995.3s$.
  & $-$\\
  \bottomrule
  \end{tabular}
  \normalsize
  \caption{Results of the runs of the chaotic orbit with $\rho=0.7$ for
 different integration schemes.}
 \label{tab-07} 
 \end{table*} 

 The initial condition $\rho=0.7$ corresponds to an orbit evolving in a strongly
 chaotic region (right panel of Fig. \ref{FigOr}). The appearing randomness of the
 motion in a strongly chaotic region produces a Fourier spectrum which corresponds
 to noise\footnote{In a Fourier spectrum of a weakly chaotic orbit we can ``detect''
 frequencies from neighboring regular orbits embedded in noise. However, such
 frequencies disappear as the chaos gets stronger, and finally we get only noise,
 see e.g. \cite{Contop02} and references therein.} (Fig. $\ref{fig-07-spec}$). In
 this case the calculations become much more computationally expensive than in the
 two other cases. Therefore, we restrict the propagation interval to
 $\tau\in[0,50000]$. The runs using the well established integrators were aborted
 very soon due to low accuracy, or in the case of CCM because the variable step
 size indeed became negative (table $\ref{tab-07}$)!
 
\begin{figure} [htp]
  \centering
  \includegraphics[width=0.4\textwidth]{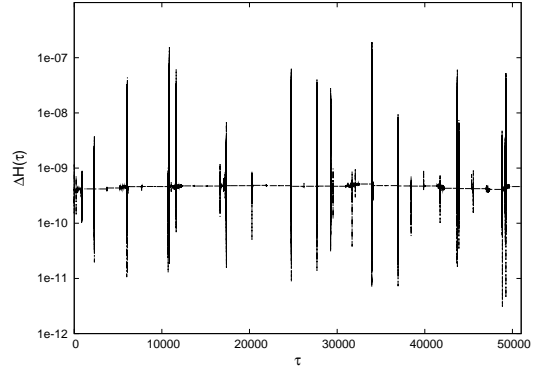}
  \caption{The relative error $\Delta H$ for the IGEM method as a function of
   proper time $\tau$, semi-logarithmic scale, for the chaotic orbit with
   $\rho=0.7$.}
  \label{fig-07-energy-IGEM}
 \end{figure}

 The only integrator to pass this test was IGEM. For IGEM, we plot the
 relative error $\Delta H$ in Fig. $\ref{fig-07-energy-IGEM}$ and observe that
 some peaks appear. This happens because, during each step, we have to solve the
 iteration $\eqref{eqn-FP-IGEM}$. For this to be a contraction, we need
 \begin{align}
  \left\|D\left(\frac{\epsilon(A\otimes I)F(\vec Z)}
  {\frac 1 2\|Df(\mathbf{Z}_1)+Df(\mathbf{Z}_s)\|}\right)\right\|\stackrel ! <1~~,
 \end{align}
  with
 \begin{align}
  D\left(\hdots\right)_{ij}
  &=\frac{\left(\epsilon(A\otimes I)DF(\vec Z)\right)_{ij}}
  {\frac 1 2\|Df(\mathbf{Z}_1)+Df(\mathbf{Z}_s)\|} \nonumber \\
  &-\frac{\left(\epsilon(A\otimes I)F(\vec Z)\right)_{i}}{\frac 1 2\
  |Df(\mathbf{Z}_1)+Df(
  \mathbf{Z}_s)\|^3}  \nonumber \\
  &\times \sum_{kl}\left(Df(\mathbf{Z}_1)_{kl}+Df(\mathbf{Z}_s)_{kl}\right)
  \nonumber \\
  &\times \pd{\left(Df(\mathbf{Z}_1)_{kl}
  +Df(\mathbf{Z}_s)_{kl}\right)}{\vec Z_j} ~~.
 \end{align}

 \begin{figure} [htp]
  \centering
  \includegraphics[width=0.4\textwidth]{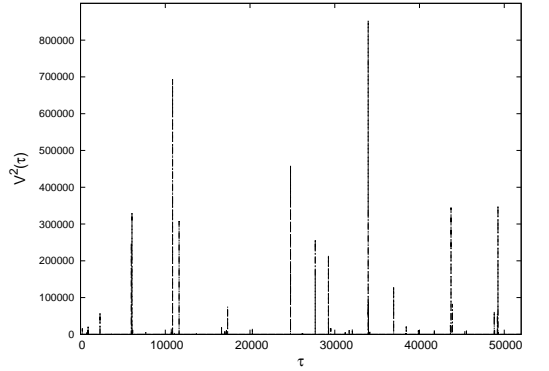}
  \caption{The 'velocity' $V(\tau)^2$ as function of proper time $\tau$, for the
  chaotic orbit with $\rho=0.7$.}
  \label{fig-07-velocity}
 \end{figure} 

 The norm of the first term is smaller than $1$ by construction of the algorithm
 but the second term can be large. This happens at points, where $f(\mathbf y)$
 becomes almost singular. To see when such a situation occurs, we plot the
 `velocity` $V^2(\tau)=\dot z^2+\dot\rho^2$ in Fig. $\ref{fig-07-velocity}$. 
 We see that at every instant of time where the Hamiltonian deviates from its
 otherwise constant value, there is a peak in the velocity. To analyze this further,
 we notice that the first peak in $V^2$ occurs at $\tau=2296.39$. We thus analyze
 the trajectory for $\tau\in[2286,2306]$. To do so, we compare the trajectory
 obtained via IGEM with $\epsilon=10^{-1}$ and once for a more 'accurate'
 calculation. In order to do the latter, we propagate the system with IGEM until
 $\tau=2296$. Then, we reduce the underlying step size $\epsilon$ to
 $\epsilon=10^{-4}$ and propagate the system until $\tau=2297$. Afterwards, we
 relax again to $\epsilon=0.1$ and calculate the configuration for
 $\tau\in[2297,2306]$. We observe that the jump in the Hamiltonian coincides with
 an inflexion point of the trajectory (i.e. the test-particle behaves like a ball
 that is thrown upon a wall), where $f(\mathbf y)$ becomes almost singular. At this
 point, IGEM with $\epsilon=10^{-1}$ cannot dissolve the trajectory with the same
 accuracy as for the other points. But, we see clearly that after the inflexion,
 the trajectory obtained via IGEM tends to follow the more accurate calculation.
 Namely, by calculating the quantity
%
% \begin{align}
 $   \mathrm{Dif}^2 =\frac{(\rho_{\text{IGEM}}-\rho_{\text{ac}})^2}{\rho_{\text{ac}}^2}
  +\frac{(\dot{\rho}_{\text{IGEM}}-\dot{\rho}_{\text{ac}})^2}{\dot{\rho}_{\text{ac}}^2}
 +\frac{(z_{\text{IGEM}}-z_{\text{ac}})^2}{z_{\text{ac}}^2}+\frac{(\dot{z}_{\text{IGEM}}-
   \dot{z}_{\text{ac}})^2}{\dot{z}_{\text{ac}}^2}~$,
% \end{align}
%
 we observe, that $\mathrm{Dif}$ is larger than $10^{-2}$ for times near the
 inflexion but decreases to below $10^{-6}$ as $\tau$ increases until $\tau=2306$.
 This observation is confirmed by Fig. $\ref{fig-07-energy-IGEM}$,  where we see
 that after the inflexion, the Hamiltonian recovers its previous value. Hence, the
 relative error $\Delta H$ jump is negligible compared to the time scale of the
 studied phenomenon.   

 To improve the performance of IGEM even more, one can hope to erase the 
 convergence problems at the singular points by using the Newton-Raphson-method to
 solve the implicit eq. (\ref{eqn-impl-IGEM}). Doing so, we notice, however,
 that the calculation time increases considerably to $T_{\text{calc}}=7521.0~s$.
 This happens, because for each iteration, the costly derivative of the rhs of eq.
 (\ref{eqn-FP-IGEM}) has to be calculated, while the average number of iterations
 per step only decreases from $3.94$ for the Fixed-Point iteration to $3.57$ for
 Newton-Raphson iteration. Therefore, even though in theory the Newton-Raphson has
 a quadratic convergence rate, it cannot catch up with the Fixed-Point iteration.  

 Our last numerical analysis argument is that IGEM is efficient and accurate in
 calculating Poincar\'e-sections because of the collocation-property. It is known
 from  Numerical Analysis that the solution at point $\mathbf y_{n+1}$ calculated
 with an $s$-stage Gauss-collocation method coincides with $U(h)$, where $ U(\tau)$
 is the interpolation polynomial through the points $(0,\mathbf y_n)$ and
 $(c_1,\mathbf y_n+\mathbf Z_1)...(c_s,\mathbf y_n+\mathbf Z_s)$. 
 Numerical Analysis further shows that the interpolation-polynomial $ U(\tau)$
 stays $\mathcal O(h^s)$ close to the real solution within the whole interval
 between $\mathbf y_n$ and $\mathbf y_{n+1}$. Thus, if we search for the root of
 $ U(\tau)$ (or rather the root of the component of the vector-valued $U(\tau)$
 which corresponds to $z$) with a fast bisection method, we get the location of the
 Poincar\'{e} section $\mathcal O(h^s)$ close to the real section. We use this
 advantage of the IGEM to enter the discussion about the appearance of chaos in the
 MSM model.

 \begin{figure} [htp]
  \centering
  \includegraphics[width=0.4\textwidth]{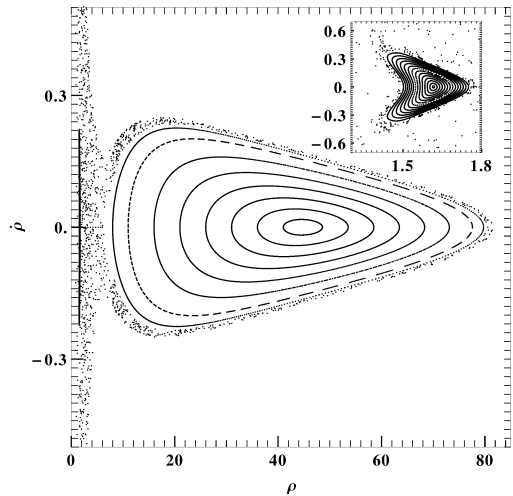}
  \caption{The Poincar\'{e} section $z=0$ for $m=2.904$, $a=1.549$, $q=0$, $\mu=0$
  and $b=0.8$, where $E=0.971$ and $L_z=9.3$. The embedded plot is a detail of the
  Poincar\'{e} section.}
 \label{figSS}
 \end{figure} 
 
 In \cite{Han08}, Han showed that chaos appears in both oblate and prolate
 deformations (with respect to the Kerr metric) of the MSM metric. However, he
 states that ``it is impossible to present chaos regions for the oblate
 deformation massive bodies, because those regions found in \cite{Dubeibe07} are
 actually inside a neutron star''. Even though Han discusses the oblate case
 $m=2.904$, $a=1.549$, $q=0$, $\mu=0$, $b=0.8$, $E=0.971$, $L_z=9.3$, he doesn't
 show the respective Poincar\'{e} section. This Poincar\'{e} section is shown in
 Fig. \ref{figSS} and it is obvious that the main island of stability, which
 lies far from the central object, is embedded in a pronounced chaotic layer. Thus,
 Han's statement should rather be that ``for realistic astrophysical objects that
 are oblate'', there are chaotic orbits of ``test particles around single oblate 
 deformation neutron stars described by'' the MSM model. However, orbits starting
 from the external chaotic layer follow trajectories like those shown in the right
 panel of Fig. \ref{FigOr}, thus these chaotic orbits in the relativistic case
 would probably plunge to the central compact object. 

 \section{Conclusion}\label{sec-VI}
 
 Applying several well established standard integration schemes to the system of
 differential equations describing geodesic motion in an example of a non-Kerr
 spacetime background (i.e. MSM \cite{MSM}), we showed that these integrators
 cannot guarantee satisfactory long term behavior for orbits in these
 non-integrable systems. Therefore, we introduce a new integration scheme
 appropriate for evolving orbits of such systems. 

 The new integration scheme effectively conserves the integrals of motion that
 such Hamiltonian systems possess, and it is well-behaved in the case of long
 term evolution of strongly chaotic orbits. Thus, the new integration scheme is
 well-suited for studying geodesic orbits in the case of spacetime backgrounds
 which are non-integrable perturbations of the Kerr spacetime. 

 Moreover, we show that chaos can appear in oblate deformations of the MSM metric
 modeling the exterior of a neutron star.

\begin{acknowledgments}
 We would like to thank B.~Bruegmann and Ch.~Lubich for useful discussions and
 suggestions. This work was supported by the DFG grant SFB/Transregio 7.
\end{acknowledgments}

\end{document}